# Theoretical insight into the interaction of $O_2$ and $H_2O$ molecules with the perfect and the defective InSe monolayers

Dongwei Ma[1,*], Tingxian Li[1], Di Yuan[1], Chaozheng He[2,*], Zhiwen Lu[2], Zhansheng Lu[3], Zongxian Yang[3], and Yuanxu Wang[1,4,*]

[1]*School of Physics, Anyang Normal University, Anyang 455000, China*
[2]*Physics and Electronic Engineering College, Nanyang Normal University, Nanyang 473061, China*
[3]*College of Physics and Materials Science, Henan Normal University, Xinxiang 453007, China*
[4]*Institute for Computational Materials Science, School of Physics and Electronics, Henan University, Kaifeng 475004, China*

**Abstract**

By using first-principles calculation, the interaction of $O_2$ and $H_2O$ molecules with the pristine and the defective InSe monolayers is studied. It is predicted that the single Se and In vacancies exhibit significantly enhanced chemical activity toward the adsorbates compared with the perfect InSe lattice site, and the Se vacancies have a much higher chemical activity than the In vacancies. $H_2O$ molecule should be only physisorbed on the various InSe monolayers at ambient conditions, according to the calculated energies. The doping of the various InSe monolayers is discussed by the physisorbed $H_2O$. The vacancies show a much higher chemical activity toward $O_2$ than $H_2O$. Although $O_2$ molecules are still physisorbed on the pristine InSe monolayer, they will be chemisorbed on the defective InSe monolayers. Especially, our calculated energies suggest that the surface oxidation of the 2D InSe semiconductor should be dominated by the defects that expose under-coordinated host atoms, especially In atoms. Our theoretical results can help better understanding the doping and the oxidation of the 2D InSe semiconductor under ambient conditions.

*Corresponding author. E-mail: dwmachina@126.com (Dongwei Ma).
*Corresponding author. E-mail: hecz2013@nynu.edu.cn (Chaozheng He).
*Corresponding author. E-mail: wangyx@henu.edu.cn (Yuanxu Wang).

# 1. Introduction

To overcome the scaling-limit of the Si-based devices, in recent years, researchers have paid great attention to explore the application of two-dimensional (2D) layered semiconducting materials in modern semiconductor engineering.[1] As one of the typical III-VI semiconductors, bulk indium selenide (InSe) has a layered crystal structure. The bulk γ-InSe phase has a direct band gap of 1.26 eV. However, due to the quantum confinement effect, 2D InSe crystal transforms from a direct to an indirect bandgap semiconductor as the number of layers is reduced.[2, 3] Consequently, the InSe monolayer has an indirect band gap of 2.11 eV.[4-6] Experimentally, 2D InSe thin films can be synthesized by using the mechanical exfoliation, liquid exfoliation, and pulsed laser deposition methods.[3, 6-8] It is demonstrated that the produced 2D InSe thin films are promising 2D semiconducting materials for high-performance electronic and optoelectronic devices.[9]

2D electron gas induced by the field effect at the surface of 2D InSe thin films has low-temperature mobility on the order of $10^3$ cm$^2$ V$^{-1}$ S$^{-1}$.[10, 11] Bandurin et al. reported that the electron mobility of the few-layer InSe encapsulated in the hexagonal boron nitride under an inert atmosphere is measured to be $10^3$ and $10^4$ cm$^2$ V$^{-1}$ S$^{-1}$ at room temperature and 4 K, respectively.[5] The values of electron effective mass are (0.14±0.01) $m_0$ for six-layer InSe and (0.17±0.02) $m_0$ for three-layer InSe,[5] much lighter than that of the monolayer MoS$_2$ ($m^* = 0.45\ m_0$) [12]. These reports suggest that 2D InSe hold great promise for fabricating ultrathin-body high-mobility nanoelectronic devices. Moreover, few-layer InSe photodetector shows a broadband photodetection with high responsivity[4, 6, 13, 14] and a gate-tunable behavior[15], making promising application in photodetection. On the theoretical aspect, the basic structural and electronic properties of the monolayer and few-layer InSe have been investigated based on density functional theory (DFT) calculations.[16-18] It is shown that an electric field perpendicular to the InSe monolayer can induce a semiconductor to metal transition.[17] The element doping effect on the electronic and magnetic properties of the InSe monolayer has also been studied very recently.[19]

One of the most important features of the 2D material is its inherent large surface-to-volume ratio and high ratios of exposed surface atoms, which is highly favorable for the gas molecule adsorption and the surface-related application. When exposed to air, it is unavoidable that the property of 2D materials and the device performance built with these materials will be influenced due to the interaction with $H_2O$ and $O_2$.[20-25] For example, Zhang et al. experimentally demonstrated that a bandgap of ~ 0.111 eV can be achieved in bilayer graphene by the *p*-doping from $H_2O/O_2$ of the bottom layer and the *n*-doping from the deposited triazine of the top layer.[20] Late et al. reported that $H_2O$ adsorption can lead to hysteresis in the $MoS_2$ monolayer field-effect transistors.[22] The influence of $H_2O$ on the optical properties of the $MoS_2$ monolayer has been investigated. It is found that the photoluminescence of the $MoS_2$ monolayer can be strongly quenched by a monolayer of sandwiched $H_2O$, which serves as *n*-type dopants.[23] $O_2$ and $H_2O$ with or without light also play a critical role in the degradation of the 2D materials in the air.[26-31] For example, based on DFT calculations and molecular dynamics simulations, the ambient degradation of black phosphorus has been ascribed to light-induced production of superoxide, its dissociation and eventual breakdown under the action of $H_2O$.[29]

Considering the importance of ambient $O_2$ and $H_2O$ in affecting the material properties, it is necessary to have an atomic level understanding on the interaction of $O_2$ and $H_2O$ molecules with the surface of 2D InSe thin film, however, which is still very limited at the present time. Politano et al.[32] found an environmental *p*-doping of 2D InSe by combining experimental and theoretical characteristics, which was attributed to the decomposition of $H_2O$ molecules at Se vacancies. However, their DFT calculations reported that most physisorption systems for the considered molecules on the perfect InSe monolayer are highly endothermic, which is not true for the system dominated with weak van der Waals (vdW) interaction. Balakrishnan et al.[33] experimentally studied the oxidation of the 2D InSe thin film in the air. It is revealed that 2D InSe thin films can be chemically stable under ambient conditions over a period of several days. However, thermal-annealing in air can significantly accelerate the oxidation and convert a few surface layers of InSe into $In_2O_3$ over a very short period of time. This behavior may be due to that there are more structural defects appearing under the thermal-annealing conditions and the structural defects often exhibit an enhancement interaction with the gas molecules.[34]

Based on DFT calculations, here we theoretically study the interaction of $O_2$ and $H_2O$ molecules with the pristine and the defective InSe monolayers. Defective InSe monolayers refer to those with single Se ($V_{Se}$) or In ($V_{In}$) point vacancies. Se and In point vacancies in the InSe monolayer exclusively expose the under-coordinated dangling In and Se atoms, respectively. These under-coordinated dangling atoms are expected to exist for other more complex vacancies, such as grain boundaries and edges. More importantly, they often exhibit enhanced chemical activity towards the gas molecules. Defect-dominated chemical activity toward gas molecules has been demonstrated for various 2D materials such as graphene[35-37] and $MoS_2$[38-41], both experimentally and theoretically.

In the work, the stability, and the atomic and electronic structures of the $V_{Se}$ and $V_{In}$ systems are firstly investigated. Then we study the adsorption of $O_2$ and $H_2O$ molecules on the pristine and the defective InSe monolayers, in terms of atomic configurations, adsorption energies, and electronic structures. Finally, results on the dissociation of $O_2$ on the various InSe supports are presented. Our theoretical results can give deep insights into the role of $O_2$ and $H_2O$ in doping and the oxidation of the 2D InSe semiconductor under ambient conditions. These insights will provide guidance for experimental control and tailing the physical properties of 2D InSe semiconductors, and advance the application of this fascinating material.

## 2. Computational details

### 2.1. Basic parameters

The calculations were performed with the projected augmented wave (PAW) formalism of DFT, as implemented in Vienna Ab-initio Simulation Package (VASP).[42-44] The generalized gradient approximation (GGA) with Perdew-Burke-Ernzerhof (PBE) for the exchange-correlation energy is used.[45] Grimme's DFT-D3 method[46] is adopted to account for the vdW effects between the adsorbed molecule and the substrate. The cutoff energy for the planewave basis set is taken as 450 eV. Structure optimizations were performed until the Hellmann-Feynman force on each atom less than 0.02 eV/Å.

The convergence of the total energy is considered to be achieved until two iterated steps with energy difference less than 10$^{-5}$ eV. During the optimization, all the internal coordinates are allowed to relax with a fixed lattice constant. Furthermore, the climbing image nudged elastic band (CINEB) method is used to find the minimum-energy path (MEP) for the transition between different states.[47] The spring constants between adjacent images are set to -5.0 eV/Å$^2$.

For the adsorption of O$_2$ and H$_2$O molecules, the 4×4 supercell has been used with the lateral size of the supercell larger than 16 Å. The distance between the InSe monolayer and its neighboring image is ~ 16 Å along the vacuum direction, which is sufficiently large to avoid the interactions between them. For such supercell, a 2×2×1 Monkhorst-Pack[48] *k*-point in the Brillouin zone is sampled for the geometry optimization and 7×7×1 for the densities of states (DOS) calculations. Unless otherwise specified, the calculations were performed with the consideration of spin-polarization.

## 2.2. Defect formation energy and adsorption energy

The defect formation energy, $E_f$, is defined as follows:

$$E_f = E_{tot}(\text{defect}) - E_{tot}(\text{pristine}) + \mu_{host}, \quad (1)$$

where $E_{tot}(\text{defect})$ and $E_{tot}(\text{pristine})$ are the total energies of the defective and the pristine InSe monolayers, respectively. $\mu_{host}$ is the chemical potential of the removed Se or In atoms. With such a definition, defect with low formation energies will occur in high concentrations.[49, 50]

The values of $\mu_{host}$ in Eq. (1) and thus $E_f$ largely depend on the experimental growth conditions. As limiting cases we consider the In-rich and Se-rich conditions. In the thermodynamic equilibrium one can assume that $\mu_{InSe} = \mu_{In} + \mu_{Se}$, where $\mu_{InSe}$ is the total energy per InSe formula unit. Under In-rich conditions, the In chemical potential ($\mu^0_{In}$) is equal to the total energy per In atom in its reference phase (i.e. the In bulk having monatomic body-centered tetragonal structure[51]). Therefore, for In-rich conditions the Se chemical potential can be written as $\mu_{Se}(\text{In-rich}) = \mu_{InSe} - \mu^0_{In}$. And for Se-rich conditions, the Se chemical potential ($\mu^0_{Se}$) is equal to the total energy per Se atom in its reference phase (i. e. the Se bulk having hexagonal crystal lattice with space group no. 152[52]). Accordingly, for Se-rich conditions the In chemical potential can be written as $\mu_{In}(\text{Se-rich}) = \mu_{InSe} - \mu^0_{Se}$.

In order to characterize the adsorption stability of $O_2$ and $H_2O$ molecules on the various monolayer InSe supports, the adsorption energy, $E_{ad}$, is studied and defined as:

$$E_{ad} = E_{InSe} + E_M - E_{InSe-M}, \qquad (2)$$

where $E_M$ denotes the total energy of the free $O_2$ or $H_2O$ molecule, and $E_{InSe}$ and $E_{InSe-M}$ denote the total energies of the various monolayer InSe supports without and with the adsorbed molecules, respectively. With this definition, a positive (negative) value of the adsorption energy means that the adsorption process is exothermic (endothermic) and energetically favorable (unfavorable).

### 2.3. Charge density difference

To better understand the interaction and the electron transfer between the adsorbed molecules and the various supports, the charge density difference (CDD) is studied. The CDD is calculated by the formula $\Delta\rho = \rho_{M/InSe} - \rho_M - \rho_{InSe}$, where $\rho_{M/InSe}$, $\rho_{InSe}$, and $\rho_M$ are the total charge densities of the support with the adsorbed molecule, the bare support, and the isolated molecule, respectively. $\rho_M$, and $\rho_{InSe}$ are calculated with each component at the same position in the adsorption structure.

### 3. Results and discussion

### 3.1. Properties of the defective InSe monolayers

Firstly, the properties of the pristine InSe monolayers are investigated. As shown in Fig. 1(a), the InSe monolayer with a hexagonal symmetry structure consists of four parallel atomic planes stacked in the Se-In-In-Se sequence, where two In atomic planes are sandwiched by two Se atomic planes. The calculated lattice constant of the InSe monolayer is about 4.05 Å, in good agreement with those reported in Ref.[18, 53]. Accordingly, the Se-In bond length is 2.68 Å and the In-In bond length 2.79 Å. The total DOS (TDOS) of the adopted (4×4) supercell are shown in the upper panel of Fig. 1(b), from which the calculated band gap of 1.24 eV is close to those of Ref.[18, 19]. Note that this value is largely underestimated compared with the measured values (2.11 eV[4-6]), due to the well-known deficiency in GGA. The local DOS (LDOS) projected on the individual Se and In atoms are shown in the lower panel of Fig. 1(b). The valance band mainly consists of Se $4p$ and In $5p$ states, and the conduction band is mainly contributed by the Se 4$p$ and In 5$s$ states, which is also consistent with previous results[18, 19].

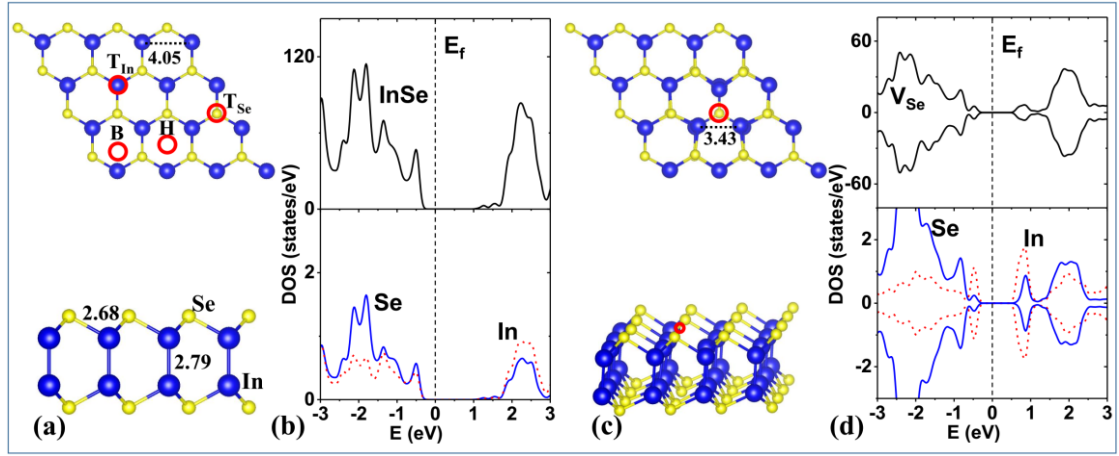

**Figure 1.** The top and side views of the (4×4) supercells of the pristine InSe monolayer and the $V_{Se}$ system are presented in (a) and (c), respectively. The Se and In atoms are denoted by the small yellow and large blue spheres, respectively. The considered adsorption sites for $O_2$ and $H_2O$ molecules are indicated in (a). The distances (in Å) between some selected neighboring atoms are given in (a) and (c). The rough position of the single Se vacancy is marked by a red circle in (c). The DOS for the pristine InSe monolayer and the $V_{Se}$ system are presented in (b) and (d), respectively, in which the TDOS are shown in the upper panels and the LDOS in the lower panels. In (b), the red and the blue lines denote the LDOS projected on the individual In and Se atoms, respectively. In (d), the red and blue lines denote the LDOS projected on the In atoms surrounding the Se vacancy and the Se atoms bonded with these In, respectively. The DOS in (b) are calculated without considering the spin-polarization. For the DOS in (d), the positive and negative values represent the spin-up and spin-down states, respectively. The $E_f$ is set to 0 eV.

The formation of the single Se vacancy is realized by removing a Se atom from the top Se atomic plane. According to Eq. (1), the calculated formation energies for the Se vacancy at the Se-poor and Se-rich limits are 0.973 and 1.893 eV, respectively. These values are comparable with the formation energy of the single S vacancy in the $MoS_2$ monolayer. For example, Liu et al. predicted that the formation energies of a S vacancy at the S-poor and S-rich limits are 0.95 and 2.35 eV, respectively.[54] The formation energy by Haldar et al. lies between about 1.3 and 2.6 eV depending on the chemical potential of the element S.[50] This suggests that the equilibrium concentrations of the Se vacancy in the InSe monolayer should be similar to those of the S vacancy in the monolayer $MoS_2$. Consequently, the presence of the Se vacancy may play an important role in the electrical transport property of 2D InSe semiconductors, similar to the role of S vacancies in the monolayer $MoS_2$ electronics.[55, 56]

The optimized structures for the $V_{Se}$ system are shown in Fig. 1(c). From the figure, it is clear that three In atoms surrounding the Se vacancy move toward to the vacancy. Three In-In bonds form with the bond lengths of 3.43 Å, which is close to the calculated In-In bond length of 3.40 Å in the bulk In. This leads to the elongation of the distance between these In atoms and their neighboring Se atoms by 0.03 Å compared with the pristine InSe monolayer. Overall, the 3-fold symmetry has been maintained around the Se vacancy, which may be confirmed in the future transmission electron microscopy image.

The TDOS for the $V_{Se}$ system are shown in the upper panel of Fig. 1(d). The features of the DOS for $V_{Se}$ system are similar to most transition metal dichalcogenides.[50] The defect states appear at the valance band edges and in the band gap towards the conduction band side. These defect states mainly come from the electronic states of the In atoms surrounding the Se vacancy with minor contribution from the Se atoms bonding with these In atoms, for which the LDOS are shown in the lower panel of Fig. 1(d). A further analysis shows that the defect state near the valance band edge mainly consists of $5p$ states of the In atoms and that in the band gap consists of 5s and 5p states of the In atoms and $4p$ states of the Se atoms. Furthermore, it is noted that the defect state in the band gap belongs to the deep acceptor state, which is detrimental to the electronic and optical properties of the 2D InSe semiconductor.

The single In vacancy is created by removing an In atom from the central In-In dimer. Depending on the structural relaxation pattern, three $V_{In}$ configurations with similar formation energies exist. The most stable one ($V_{In}^1$) presented in Fig. 2(a) has a formation energy of 1.499 eV at the In-poor limit and 2.419 eV at the In-rich limit, much smaller than those for the Mo vacancy in the $MoS_2$ monolayer. The predicted formation energies for the Mo vacancy are 4.2 and 7.0 eV, respectively, at the Mo-poor and Mo-rich limits, respectively.[50] The configurations shown in Figs. 2(b) ($V_{In}^2$) and 2(c) ($V_{In}^3$) are slightly less stable than the $V_{In}^1$ configuration by 0.020 and 0.021 eV, respectively. Rich reconstruction of the simple point vacancy in the 2D materials has been demonstrated for phosphorene containing a single P vacancy,[57, 58] where four energetically inequivalent states have been theoretically identified.

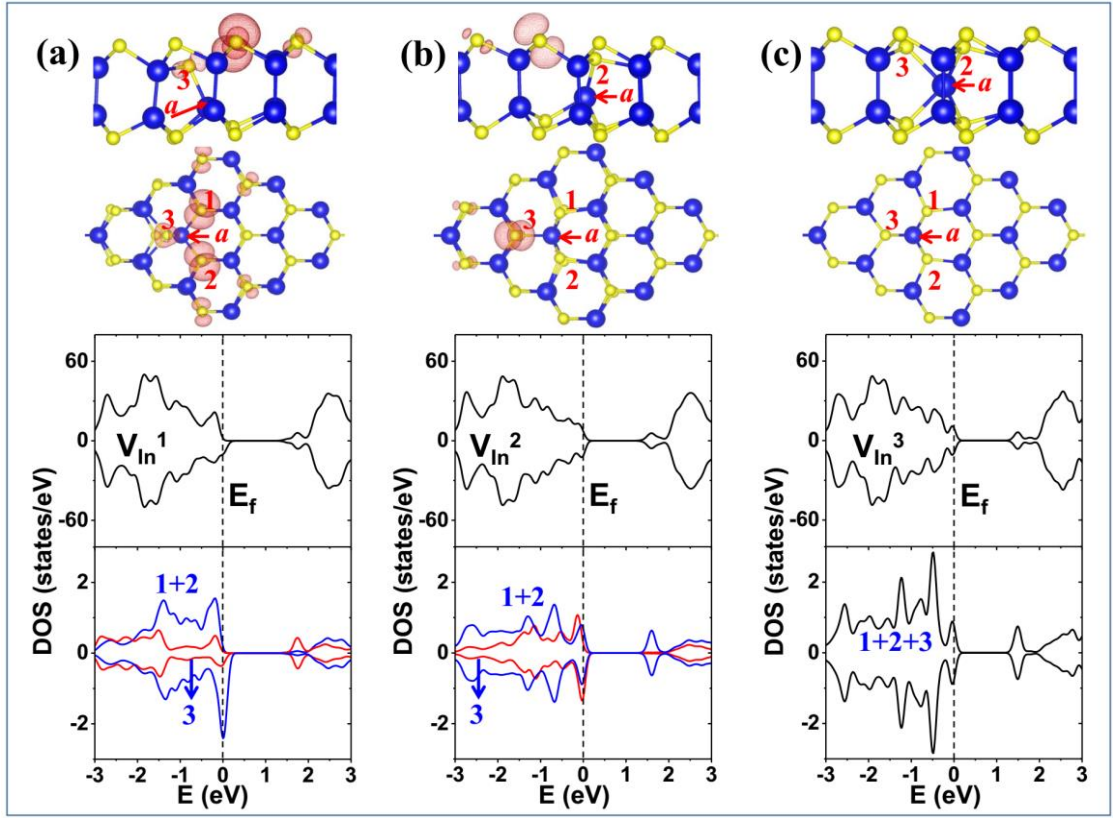

**Figure 2.** The atomic structures, the spin-density isosurfaces, and the DOS of the $V_{In}^1$, $V_{In}^2$, and $V_{In}^3$ systems are shown in (a), (b), and (c), respectively. The four atoms surrounding the single In vacancy are marked. The value of the spin-density isosurface is taken as $8 \times 10^{-4}$ e/bohr$^3$. The TDOS and the LDOS projected on the three Se atoms surrounding the In vacancy are presented in the upper and the lower panels of the DOS graph, respectively.

There are four atoms surrounding each In atom in the pristine InSe monolayer (see Fig. 1(a)). Therefore, removing an In atom from the upper In atomic plane, there will be four under-coordinated atoms surrounding the In vacancy, i. e. three Se atoms (atoms

1, 2 and 3 in Fig. 2) and one In atom (atom *a* in Fig. 2). For the $V_{In}^1$ configuration (Fig. 2(a)), Se1 and Se2 atoms are coordinated with two In atoms in the upper In atomic plane. Se3 atom moves down and In*a* atom moves up with respect to the respective atomic plane. Consequently, Se3 atom is still coordinated with three In atom with the Se3-In*a* bond length of 2.70 Å. This relaxation pattern leads to the appearance of spin-unpaired electron localized mainly on Se1 and Se2 atoms, the spin-density isosurface for which is presented in Fig. 2(a). The total magnetic moment of the whole supercell is 0.95 $\mu_B$, of which 63% is localized on the dangling Se1 and Se2 atoms. In contrast, for the $V_{In}^2$ configuration (Fig. 2(b)), Se3 atom is coordinated with two In atoms, while Se1 and Se2 atoms moving down are coordinated with three In atoms. The resulting Se1(Se2)-In*a* bond length is 2.85 Å. With this relaxation pattern, a spin magnetic moment of 0.38 $\mu_B$ has been produced, of which nearly half is localized on the dangling Se3 atom. The $V_{In}^3$ configuration is a nonmagnetic state. As presented in Fig. 2(c), In*a* atom moves up significantly, and consequently Se atoms on both sides are all bonded with this In atom, with each bond length equal to 2.90 Å.

From the TDOS in the upper panel of the DOS graph (Fig. 2), the overall feature for the three systems is that the InSe monolayers are *p*-doped due to the presence of the cation vacancy. Compared with the TDOS of the pristine InSe monolayer shown in Fig. 1(b), it can be seen that there are two defect states. One defect state crosses the Fermi level ($E_f$), while another defect state locates near the edge of the conduction band. The PDOS of the Se atoms surrounding the In vacancy are shown in the lower panel of the DOS graph. Further analyses indicate that the defect state crossing the $E_f$ mainly comes from the 4*p* states of certain Se atoms surrounding the In vacancy, depending on the specific structural relaxation pattern. For all the three systems, the defect state near the conduction band are mainly due to the hybridization between the 5*s* state of the In*a* atom and the 4*p* states of its bonded Se atoms. Due to the lower formation energy of the In vacancy (~ 1.5 eV) at the Se-rich condition than that of the Se vacancy (~ 1.9 eV), the *p*-type device based on 2D InSe with intrinsic defects may be realized under proper experimental condition.

**3.2. Adsorption of O₂ and H₂O molecules on the pristine InSe monolayer**

The adsorption of $O_2$ and $H_2O$ molecules is studied next on the pristine InSe monolayer, including the physisorption (PS), molecular chemisorption (MC), and dissociative chemisorption (DC) states. For the PS and MC states, as shown in Fig. 1(a), four adsorption sites have been considered, including the top of a Se atom ($T_{Se}$), the top of an In atom ($T_{In}$), the center of a Se-In hexagon (H), and the center of a Se-Se bridge (B). For each adsorption site, different molecule orientations have been investigated. For $O_2$ molecule adsorption, two kinds of initial configurations have been considered with the molecule being aligned either parallel or perpendicular to the surface of the InSe monolayer. For $H_2O$ molecule adsorption, the configurations with both H atom, one of two H atoms or the O atom pointing down have been investigated. After structural optimization, it is found that for both $O_2$ and $H_2O$ molecules, there are no MC states. The most stable PS states for each adsorption site and the DC states with higher stability are shown in Fig. 3 for $O_2$ and $H_2O$ molecules.

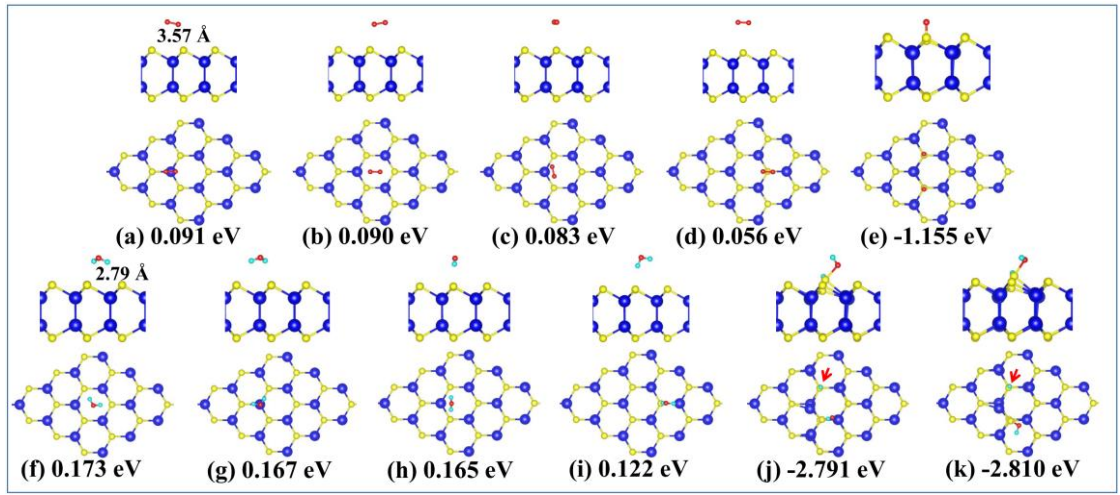

**Figure 3.** The top and side views of the possible stable adsorption configurations of $O_2$ and $H_2O$ molecules on the pristine InSe monolayer. (a)-(d) the PS states of $O_2$, (e) the DC state of $O_2$ (InSe-DC1($O_2$)), (f)-(i) the PS states of $H_2O$, and (j) and (k) the DC states of $H_2O$. The adsorption energies are given. For the most stable PS states of $O_2$ (a) (InSe-PS1($O_2$)) and $H_2O$ (f) (InSe-PS1($H_2O$)), the nearest distance between the adsorbed molecules and the surfaces are 3.57 and 2.79 Å, respectively.

As presented in Figs. 3(a)-3(d), for the PS states of $O_2$ on the pristine InSe monolayer, the adsorption energies range from 0.091 to 0.056 eV, which are much larger than the DC state (InSe-DC1($O_2$), Fig. 3(e)). This configuration is endothermic by 1.155 eV, suggesting that the dissociation of $O_2$ molecules is highly energetically unfavorable. The dissociation of $H_2O$ molecules on the pristine InSe monolayer is more energetically unfavorable than that of $O_2$ molecules. From Figs. 3(f)-3(i), the adsorption energies for the PS states of $H_2O$ molecules on the pristine InSe monolayer are between 0.173 and 0.122 eV, showing a stronger interaction with the substrate than $O_2$. Two DC states have been obtained presented in Figs. 3(j) and 3(k). According to adsorption energies, the DC states for the $H_2O$ molecule are significantly less stable than the PS states by ~ 3.0 eV. Much stronger interaction between $H_2O$ and $O_2$ with the InSe monolayer for the PS states than the DC states means that the pristine InSe monolayer will be doped by the physisorbed $H_2O$ or $O_2$ molecules by the means of surface charge transfer.[59]

The most stable PS state for the $O_2$ molecule on the InSe monolayer is shown in Fig. 3(a) (InSe-PS1($O_2$)) and has an adsorption energy of 0.091 eV. In this configuration, the adsorbed $O_2$ molecule is almost parallel to the InSe surface with the molecule siting above the $T_{In}$ site. The nearest distance between the $O_2$ molecule and the atoms of the InSe monolayer is 3.57 Å and the O-O bond length of the adsorbed $O_2$ is basically unchanged. The similar configuration has been found for the $O_2$ adsorption on the $MoS_2$ monolayer.[60] The CDD for InSe-PS1($O_2$) is presented in Fig. 4(a). It can be seen that there is electron accumulation on the $O_2$ molecule with the electron depletion on the three Se atoms under the $O_2$ molecule. The charge transfer trend is in good agreement with the result of Bader charge analysis,[61] which shows that the $O_2$ molecule gains 0.02 $e$ electrons from the pristine InSe monolayer. These results indicate that the $O_2$ molecule acts as an acceptor for the pristine InSe monolayer. The TDOS and LDOS for InSe-PS1($O_2$) are shown in Fig. 4(b). From the TDOS in the upper panel of the DOS graph, the most obvious change for the electron states near the $E_f$ is that there is a sharp impurity state in the band gap. From the LDOS shown in the lower panel, this impurity state comes from the lowest unoccupied molecular orbital of the adsorbed $O_2$ ($2\pi^*$). The in-gap state may trigger novel optical properties of the materials. From the LDOS, it is clear that no orbital hybridization between the adsorbed molecule and its underlying three atoms, further confirming the PS nature for the system. The CDD and the DOS for the configurations shown in Figs. 3(b), 3(c) and, 3(d) are presented in Figs. S1(a), S1(b), and S1(c), respectively. The DOS of these three configurations are similar to those displayed in Fig. 4(a). However, the CDD and Bader charge analysis show that the $O_2$ molecule above the $T_{Se}$ site (Fig. S1(c)) acts as a weak donor by denoting electron (0.004 $e$) to the pristine InSe monolayer. Overall, due to much smaller adsorption energy for this configuration compared with other three ones, the pristine InSe monolayer will be $p$-dope by $O_2$ molecules at ambient conditions.

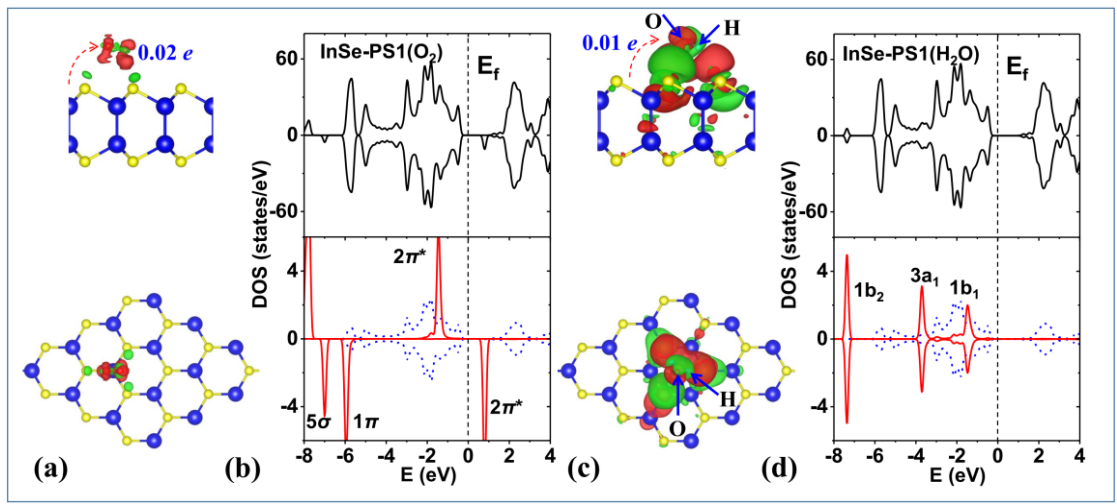

**Figure 4.** The CDD of InSe-PS1($O_2$) and InSe-PS1($H_2O$) are shown in (a) and (c), respectively. The corresponding DOS are shown in (b) and (d). For the CDD, the red and green regions represent the electron accumulation and depletion, respectively, and the value of the charge isosurface is taken as $1\times10^{-4}$ e/bohr$^3$. In (b) and (d), the solid red and the dashed blue lines denote the LDOS projected on the adsorbed molecule and its neighboring Se atoms, respectively. The rough positions of the molecular orbitals of the adsorbed molecule are indicated.

The most stable PS state for $H_2O$ molecules on the pristine InSe monolayer is presented in Fig. 3(f) (InSe-PS1($H_2O$)). This configuration has an adsorption energy of 0.173 eV. The $H_2O$ molecule sits above the H site and the nearest distance between the atoms of the adsorbed molecule and those of the support is 2.79 Å. InSe-PS1($H_2O$) has the H atoms of the $H_2O$ pointing toward the underlying Se atoms, and so do other three configurations (Figs. 3(g)-3(i)). The CDD for InSe-PS1($H_2O$) is presented in Fig. 4(c), which shows a significant charge redistribution compared with the case of $O_2$ adsorption. From the figure, it can be seen that, electron depletion occurs between the O atom of the adsorbed $H_2O$ and the support, while there is electron accumulation between the H atoms and the support. Overall, according to Bader charge analysis, the $H_2O$ molecule obtains electrons from the pristine InSe monolayer by 0.01$e$ and thus the pristine InSe monolayer is *p*-doped by the adsorbed $H_2O$ molecule. The similar doping behavior has been found for $H_2O$ adsorption on the $MoS_2$ monolayer[60] and germanene[62]. From the TDOS shown in the upper panel of Fig. 4(d), there are no obvious changes for the electron states near the $E_f$ compared with those of the pristine InSe monolayer. The LDOS in the lower panel show that only the highest occupied molecular orbital ($1b_1$ state) is slightly broadened, while the $1b_2$ and $3a_1$ states remain unperturbed. The CDD and the DOS for the configurations shown in Figs. 3(g), 3(h) and, 3(i) are presented in Figs. S2(a), S2(b), and S2(c), respectively. Similar charge transfer behaviors and electronic structures are observed for these three configurations compared with the most stable one. Therefore, the pristine InSe monolayer will be also *p*-doped by the adsorbed $H_2O$ molecules at ambient conditions.

### 3.3. Adsorption of $O_2$ and $H_2O$ molecules on $V_{In}$

Various adsorption states have been considered for the adsorption of $O_2$ and $H_2O$ molecules on the $V_{In}^1$, $V_{In}^2$ and $V_{In}^3$ surfaces. Similar to the pristine InSe monolayer, there are only PS and DC states existing. For $O_2$ adsorption on the $V_{In}^1$ surface, the selected PS state ($V_{In}^1$-PS1($O_2$)) is shown in Fig. 5(a). For this configuration, the adsorbed $O_2$ molecule is almost parallel to the support plane and sits between Se1 and Se2 atoms. The calculated nearest distance between the atoms of the adsorbed $O_2$ and the support is 3.45 Å and the adsorption energy is 0.069 eV, indicative of typical PS state. The selected PS states for $O_2$ on $V_{In}^2$ and $V_{In}^3$ are shown in Figs. S3(a) and S3(b), respectively, both of which have an adsorption energy of slightly larger than 0.1 eV. For $H_2O$ adsorption on $V_{In}^1$, two selected PS configurations with higher stability are shown in Figs. 5(d) ($V_{In}^1$-PS1($H_2O$)) and 5(e) ($V_{In}^1$-PS2($H_2O$)). Both configurations have the H atoms of the adsorbed $H_2O$ pointing toward the surface, similar to the cases of $H_2O$ adsorption on the pristine InSe monolayer. The adsorption energies for $V_{In}^1$-PS1($H_2O$) and $V_{In}^1$-PS2($H_2O$) are 0.245 and 0.186 eV, respectively, larger than those for $H_2O$ on the pristine InSe monolayer. The selected PS states for $H_2O$ on the $V_{In}^2$ and $V_{In}^3$ surfaces are shown in Figs. S4(a)-S4(d), which also show higher adsorption stability than $H_2O$ adsorption on the pristine InSe monolayer.

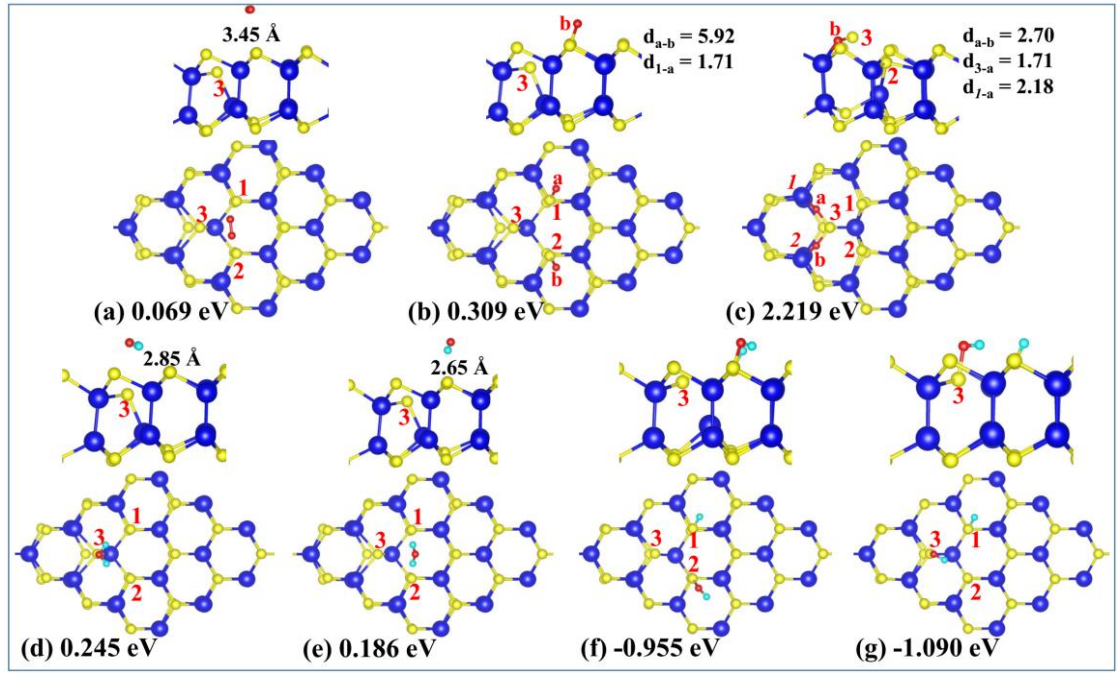

**Figure 5.** The top and side views of the stable PS configurations for $O_2$ and $H_2O$ molecules on $V_{In}^1$ and the selected DC states. (a) the PS state of $O_2$, and (d) and (e) the PS states of $H_2O$. (b) and (c) the DC states of $O_2$, and (f) and (g) the DC states of $H_2O$. Some selected structural parameters (in Å) are given for $V_{In}$-DC1($O_2$) and $V_{In}$-DC2($O_2$), shown in (b) and (c), respectively. Note that, to obtain the possible DC states for the molecules on $V_{In}$, all the three $V_{In}$ systems, i.e. $V_{In}^1$, $V_{In}^2$ and $V_{In}^3$, are investigated. And the adsorption energies for the DC states are calculated with respect to the free molecule and the $V_{In}^1$ configuration, irrespective to the initial configurations. The nearest distances between the adsorbed molecules and the surfaces for $V_{In}^1$-PS1($O_2$) (a), $V_{In}^1$-PS1($H_2O$) (d), and $V_{In}^1$-PS2($H_2O$) (e) are 3.45, 2.85 and 2.65 Å, respectively.

Further study on the chemisorption reveals a significantly enhanced chemical activity of the In vacancy towards $O_2$ and $H_2O$ molecules. For $O_2$ adsorption, from various initial configurations, four DC states have been obtained. The adsorption energies of the DC states presented in Figs. 5(b) ($V_{In}$-DC1($O_2$)) and 5(c) ($V_{In}$-DC2($O_2$)) are 0.309 and 2.219 eV, respectively, and those presented in Figs. S3(c) and S3(d) are about 0.1 and -0.2 eV, respectively. These values are much larger than that that (-1.155 eV) of InSe-DC1($O_2$). For $H_2O$ adsorption on the $V_{In}$ surface, five DC states have been obtained, the adsorption energies for which range from -0.955 to -1.090 eV. The most and least stable states are presented in Figs. 5(f) and 5(g), respectively, and other three configurations are shown in Figs. S4(e)-S4(g). Compared with the DC states (with adsorption energies ~ -2.8 eV) for $H_2O$ on the pristine InSe monolayer, the $V_{In}$ surface also shows a significantly enhanced chemical activity toward $H_2O$ molecules.

In the following, it is necessary to have further insights into the properties of the configurations of higher stability. From above, it is noted that, for $O_2$ adsorption on the $V_{In}$ surface, the DC states $V_{In}$-DC1($O_2$) and $V_{In}$-DC1($O_2$) presented in Figs. 5(b) and 5(c), respectively, are much more stable than the PS states and other DC states. Structural parameters for both configurations can be seen in the figures. Rather large O-O distances of 5.92 Å and 2.70 Å in Figs. 5(b) and 5(c), respectively, confirm the complete dissociation of the adsorbed $O_2$ on the $V_{In}$ surface. Furthermore, both $V_{In}$-DC1($O_2$) and $V_{In}$-DC2($O_2$) are magnetic with the total spin magnetic moment of 1.00 $\mu_B$ for each state and the magnetic moments are mainly localized on the O atoms and their bonded Se atoms. Bader charge analysis shows that larger than 80% magnetic moments are localized on these atoms. This picture can be further confirmed by the spin-density isosurfaces presented in Figs. 6(a) and 6(b). The TDOS and relevant LDOS for $V_{In}$-DC1($O_2$) and $V_{In}$-DC2($O_2$) are also shown in Figs. 6(a) and 6(b), respectively. From the TDOS in the upper panels of the DOS graphs, both configurations have sharp peaks around the $E_f$. From the LDOS shown in the lower panels of DOS graphs, these sharp peaks mainly come from the adsorbed O and its bonded Se atoms ($p$ states), which are responsible for the formation of the spin magnetic moment.

For $H_2O$ adsorption on the $V_{In}$ surface, the PS states are still much more stable than the DC states, similar to $H_2O$ adsorption on the pristine InSe monolayer surface. Therefore, the physisorbed $H_2O$ will play a role in modifying the electronic property of the $V_{In}$ system. Bader charge analysis shows that for $V_{In}^1$-PS1($H_2O$) and $V_{In}^1$-PS2($H_2O$) the adsorbed $H_2O$ molecules gain electrons of 0.01 $e$ and 0.02 $e$, respectively, which suggests that the $V_{In}^1$ system is $p$-doped by the adsorbed $H_2O$ molecule. The CDD and the DOS for $V_{In}^1$-PS1($H_2O$) and $V_{In}^1$-PS2($H_2O$) are shown in Figs. 6(c) and 6(d), respectively. From the CDD, there is a significant redistribution of the charge density due to $H_2O$ adsorption for $V_{In}^1$-PS1($H_2O$) compared with $V_{In}^1$-PS2($H_2O$). From the TDOS shown in the upper panel of the DOS graphs, there are no obvious changes for the electronic states near the $E_f$ compared with those of the bare support. The LDOS in the lower panel show that the $1b_1$ state is more broadened for $V_{In}^1$-PS1($H_2O$) than that for $V_{In}^1$-PS2($H_2O$). This, together with the significant charge redistribution for $V_{In}^1$-PS1($H_2O$) compared with $V_{In}^1$-PS2($H_2O$), could be responsible for the stronger adsorption of $H_2O$ molecule for the former than the latter configuration. Similar charge transfer behaviors and electronic structures are observed for $H_2O$ on the $V_{In}^2$ and $V_{In}^3$ surfaces, as shown in Fig. S5. Therefore, similar to the pristine InSe monolayer, the $V_{In}$ system will be also $p$-doped by the physisorbed $H_2O$ molecules at ambient conditions.

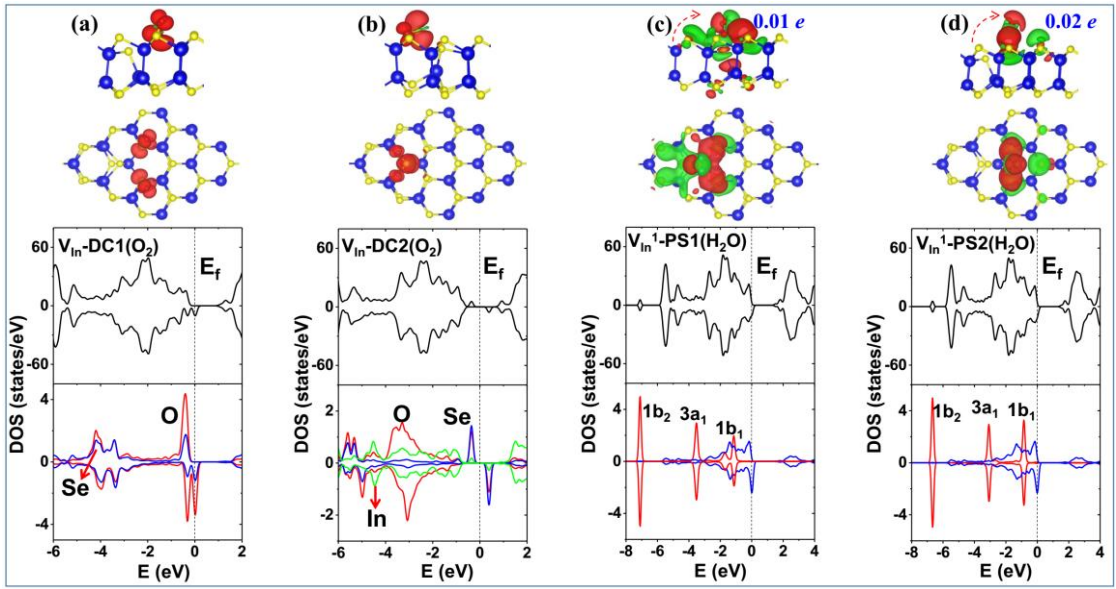

**Figure 6.** The spin-density isosurfaces and the DOS for $V_{In}$-DC1($O_2$) and $V_{In}$-DC2($O_2$) are given in (a) and (b), respectively. The CDD and the DOS for $V_{In}^1$-PS1($H_2O$) and $V_{In}^1$-PS2($H_2O$) are given in (c) and (d), respectively. The values of the charge and the spin-density isosurfaces are taken as $1\times10^{-4}$ and $1\times10^{-3}$ $e$/bohr$^3$, respectively. In (a) and (b), the LDOS in the lower panels of the DOS graphs are those of the O atoms and their bonded Se (In) atoms. In (c) and (d), the red and blue lines denote the LDOS projected on the adsorbed molecule and its neighboring Se atoms, respectively.

### 3.4. Adsorption of $O_2$ and $H_2O$ molecules on $V_{Se}$

For $O_2$ adsorption on $V_{Se}$, there are PS, MC and DC states existing. The adsorption energy for the PS state ($V_{Se}$-PS1($O_2$)) is 0.131 eV. The atomic structures for this state are shown in Fig. 7(a). The presence of the Se vacancy will significantly enhance the chemical activity of the InSe monolayer surface. As shown in Figs. 7(b) and 7(c), two MC states have the adsorption energies larger than 2 eV and the O-O bond lengths of the adsorbed $O_2$ are close to 1.5 Å. The most stable one, $V_{Se}$-MC1($O_2$) (Fig. 7(b)), has an adsorption energy of 2.354 eV, slightly more stable than $V_{Se}$-MC2($O_2$) (Fig. 7(c)) by ~ 0.1 eV. Interestingly, for $V_{Se}$-MC1($O_2$) (Fig. 7(b)) the $O_a$ atom sits between the upper Se and In atomic planes, while the $O_b$ atom nearly in the upper In atomic plane. On the contrary, for $V_{Se}$-MC2($O_2$) (Fig. 7(c)) the $O_a$ atom sits nearly in the upper In atomic plane, while the $O_b$ atom nearly in the upper Se atomic plane. Importantly, as shown in Fig. S6, the transition from the $V_{Se}$-MC2($O_2$) configuration to the configuration ($V_{Se}$-MC3($O_2$)) equivalent to $V_{Se}$-MC1($O_2$) only needs to overcome an energy barrier of ~0.01 eV. Especially, we find that the dissociation of $O_2$ in $V_{Se}$-MC2($O_2$) to form separated O atoms must first transform to $V_{Se}$-MC3($O_2$). Therefore, in the following, we only consider the possible dissociation of the adsorbed $O_2$ in $V_{Se}$-MC1($O_2$).

To further study $O_2$ dissociation in $V_{Se}$-MC1($O_2$), various DC states have been considered with one O atoms occupying the Se vacancy site and another O atom at various sites neighboring the Se vacancy. Consequently, total five DC states have been obtained. The most stable one, $V_{Se}$-DC1($O_2$), is shown in Fig. 7(d) and the other four ones shown in Figs. S7. The $V_{Se}$-DC1($O_2$) configuration has an adsorption energy of 4.696 eV, much more stable than the other four ones with the adsorption energies ranging from 2.2 to 2.5 eV. Therefore, the $O_2$ dissociation in $V_{Se}$-MC1($O_2$) should be very prone to produce the $V_{Se}$-DC1($O_2$) configuration compared with other DC states. In fact, our CINEB calculation fails to locate the MEPs for the transition from the $V_{Se}$-MC1($O_2$) configuration to the four DC states shown in Figs. S7. Therefore, the $V_{Se}$-

DC1(O$_2$) configuration is more relevant for the following discussion.

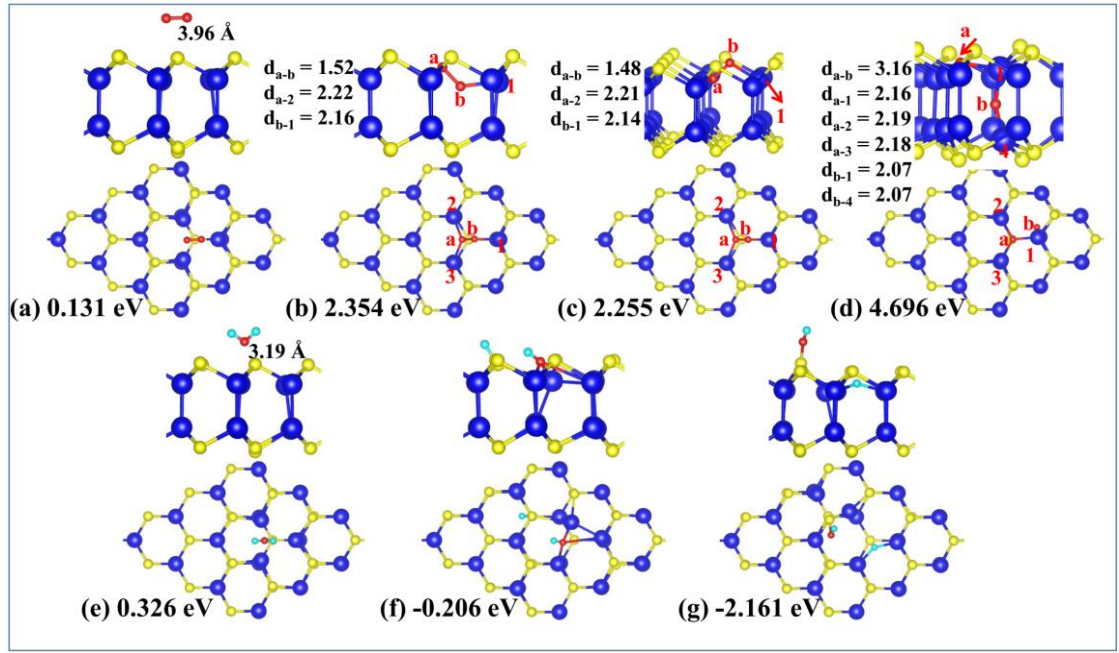

**Figure 7.** The top and side views of the PS, MC, and DC configurations for O$_2$ and H$_2$O molecules on V$_{Se}$. (a) the PS state of O$_2$, and (e) the PS states of H$_2$O. (b) and (c) the MC states of O$_2$. (d) the DC state of O$_2$, and (f) and (g) the DC states of H$_2$O. Some selected structural parameters (in Å) are given for V$_{Se}$-MC1(O$_2$), V$_{Se}$-MC2(O$_2$), and V$_{Se}$-DC1(O$_2$), shown in (b), (c), and (d), respectively. The nearest distance between the adsorbed molecules and the surfaces for V$_{Se}$-PS1(O$_2$) (a) and V$_{Se}$-PS1(H$_2$O) (e) are 3.96 and 3.19 Å, respectively.

As shown in Fig. 7(d), in the V$_{Se}$-DC1(O$_2$) configuration, the O$_a$ atom occupies the Se vacancy and the O$_b$ atom is nearly inserted between the In-In dimer. It is noted that, the O$_a$ atom lies below the upper Se atomic plane, due to the small atomic radius of O atom compared with Se atom.[63] The distance between O$_a$ and O$_b$ atoms is 3.16 Å. The bond lengths between O$_a$ and its three neighboring In atoms are all about 2.2 Å and those between O$_b$ and its two neighboring In atoms all about 2.1 Å. These structural parameters suggest that the O-O bond is completely dissociated by forming O-In bonds. The DOS for V$_{Se}$-DC1(O$_2$) state is shown in Fig. 8(b). From the TDOS in the upper panel, the system is nonmagnetic as indicated by the symmetric DOS for the spin-up and spin-down states. Importantly, there are no defect states in the band gap. Therefore, the deep acceptor states induced by the Se vacancy are removed completely by the doping of the dissociated $O_2$ molecules. Similar behavior has been observed for the GaS and $MoS_2$ monolayer.[64]

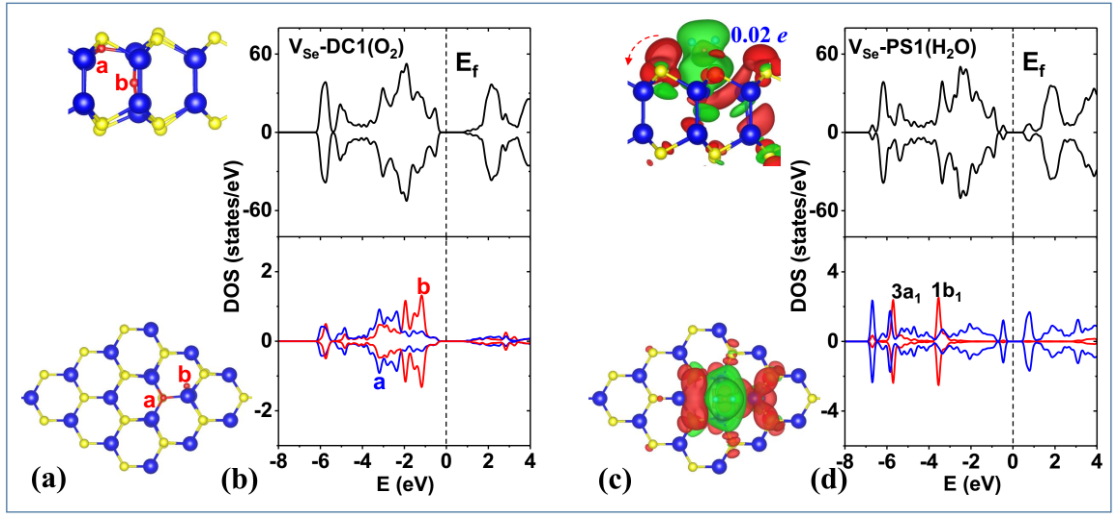

**Figure 8.** The atomic configurations and the DOS for $V_{Se}$-DC1($O_2$) are given in (a) and (b), respectively. The CDD and the DOS for $V_{Se}$-PS1($H_2O$) are given in (c) and (d), respectively. The values of the charge isosurfaces are taken as $1\times10^{-4}$ e/bohr$^3$. In (a) and (b), the LDOS in the lower panels of the DOS graphs are those of the dissociated O atoms as shown in (a). In (d), the red and blue lines denote the LDOS projected on the adsorbed molecule and the exposed three In atoms, respectively.

Various configurations have been considered for $H_2O$ adsorption above the Se vacancy. However, after structural relaxation, only one PS state (i.e. $V_{Se}$-PS1($H_2O$)) has been obtained Figs. 7(e). The adsorption energy for this state is 0.326 eV, larger than those for $H_2O$ adsorption on the other InSe surfaces in this work. The most and least stable DC states for $H_2O$ adsorption on the $V_{Se}$ surface are shown in Figs. 7(f) and 7(g), respectively. For $V_{Se}$-DC1($H_2O$), the OH group and the H atom of the dissociated $H_2O$ sit at the Se vacancy and above its one neighboring Se atom, respectively. In contrast, for $V_{Se}$-DC2($H_2O$), the OH group sits above one of Se atom neighboring the Se vacancy while the separated H atom sits at the Se vacancy. $V_{Se}$-DC1($H_2O$) has an adsorption energy of -0.206 eV, much more stable than the $V_{Se}$-DC2($H_2O$) state by ~ 2 eV. Therefore, configurations with the separated OH group occupying the Se vacancy are more favorable. Two such configurations with lower stability are shown in Fig. S8. Although the presence of Se vacancies also enhances the chemical activity of the InSe monolayer toward the $H_2O$ molecule, the adsorbed $H_2O$ will exist in the PS state at ambient conditions, due to that the PS states are still much more stable than the DC states and the DC states are endothermic.

The $V_{Se}$-PS1($H_2O$) configuration has the O atom of the adsorbed $H_2O$ pointing toward the Se vacancy and the nearest distance between the $H_2O$ and the surface is 3.19 Å. This is in sharp contrast with the PS states presented above, where one or two H atoms of the adsorbed $H_2O$ point toward the surface. This structural characteristic may result from the electrostatic attraction interaction between the negatively charged O ion and the positively charged In ions exposed to the adsorbate on the $V_{Se}$ surface. The CDD and the DOS for $V_{Se}$-PS1($H_2O$) are shown in Figs. 8(c) and 8(d), respectively. The CDD (Figs. 8(c)) shows that electron depletion occurs mainly on the $H_2O$ molecule while electron accumulation mainly on the four Se atoms around the Se vacancy. The charge transfer direction can be quantitatively validated by Bader charge analysis, which predicts that the $H_2O$ molecule denotes electrons by 0.02 $e$ to the support. Therefore, contrary to the other InSe systems, the $V_{Se}$ system can be $n$-doped by the physisorbed $H_2O$. The TDOS in the upper panel of Figs. 8(d) also show that there are no additional electronic states within the band gap compared with the bare $V_{Se}$ system.

### 3.5. $O_2$ dissociation

The present study shows that, irrespective of the presence of defects, $H_2O$ molecule can only be physisrobed on the various InSe surfaces at ambient conditions, due to that the PS states for $H_2O$ are always much more stable than the chemisorption states and the latter are always endothermic processes, as shown in Table 1. In contrast, the vacancies, especially Se vacancies, can greatly enhance the chemical activity toward the adsorbed $O_2$, as evidenced by the significantly increased adsorption energies for the chemisorption states (Table 1). Therefore, the adsorption of $O_2$ and its further dissociation on the InSe surface should play a critical role in the possible degradation of 2D InSe semiconductors in the air, which is out of the scope of the present study. Nevertheless, it is necessary to investigate the kinetic process from the physisorbed $O_2$ to the chemisorbed $O_2$, in order to gain a further insight into the effect of the exposed under-coordinated In or Se atoms on the oxidation of 2D InSe semiconductors.

**Table 1.** Adsorption energies (in eV) of the various states for $O_2$ and $H_2O$ molecules on the pristine InSe monolayer, the $V_{In}^1$, and the $V_{Se}$ systems. The values outside and inside the parentheses are for $O_2$ and $H_2O$ molecules, respectively. Note that the adsorption stability for $O_2$ and $H_2O$ molecules on $V_{In}^2$ and $V_{In}^3$ is similar to that on $V_{In}^1$.

|  | InSe | $V_{In}^1$ | $V_{Se}$ |
| --- | --- | --- | --- |
| Physisorption | 0.091 (0.173) | 0.069 (0.245) | 0.131 (0.326) |
| Molecular chemisorption |  |  | 2.354 |
| Dissociative chemisorption | -1.155 (-2.791) | 2.219 (-0.955) | 4.696 (-0.206) |

For $O_2$ on the pristine InSe monolayer, the atomic structures and the relevant energies are shown in Fig. 9 for the initial state (IS), transition state (TS), and final state (FS) along the MEP for the transition from the PS state to the DC state. It can be seen that the process is endothermic by 1.24 eV and needs to overcome an energy barrier of 2.85 eV. Thus $O_2$ dissociation on the pristine InSe monolayer is highly unfavorable, both energetically and dynamically. This indicates that 2D InSe semiconductors without any structural defects should be well resistant to oxidation and very stable under ambient conditions. The presence of the In vacancy should be able to significantly accelerate the oxidation of 2D InSe semiconductors. For $O_2$ dissociation on the $V_{In}^1$ surface (Fig. 10), the $V_{In}$-DC1($O_2$) configuration is selected as an intermediate state (MS) and the $V_{In}$-DC2($O_2$) configuration selected as the FS. The approaching of the physisorbed $O_2$ molecule to the surface is exothermic by 0.24 eV and the whole $O_2$ dissociation process exothermic by 2.15 eV. The energy barrier for the approaching of the $O_2$ molecule to the surface is 0.94 eV and that for the transition from the MS to FS is 1.03 eV. In addition, it is found that the energy barrier for the transition from the PS state to $V_{In}$-DC1($O_2$) on $V_{In}^2$ (0.98 eV) or $V_{In}^3$ (0.95 eV) is similar to that of $V_{In}^1$, as shown in Fig. S9. The Se vacancy should have a more significant effect on the oxidation of 2D InSe semiconductors in comparison with the In vacancy. As shown in Fig. 11, the transition from the $V_{Se}$-PS1($O_2$) configuration (IS) to the $V_{Se}$-MC1($O_2$) configuration

(MS) is highly exothermic by 2.22 eV and the whole $O_2$ dissociation process is exothermic by 4.57 eV. Furthermore, the highest energy barrier for the whole process is only 0.17 eV corresponding to the transition from the MC state to the DC state, much smaller than those for the pristine InSe monolayer (2.85 eV) and the $V_{In}$ system (1.03 eV).

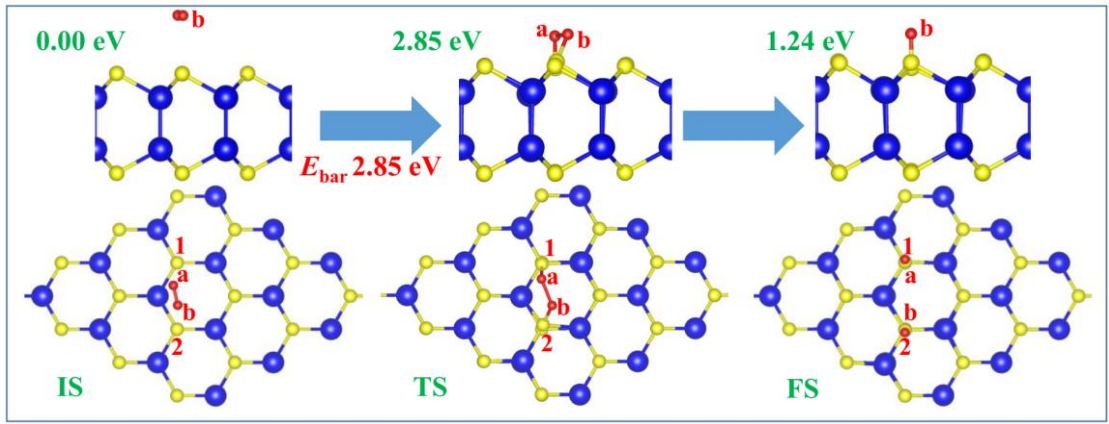

**Figure 9.** The atomic configurations of the IS, TS, and FS along the MEP for the $O_2$ dissociation on the pristine InSe monolayer. The energies of the IS, TS, and FS are given with respect to the total energy of the IS. The InSe-PS3($O_2$) (Fig. 3(c)) and InSe-DC1($O_2$) (Fig. 3(e)) configurations are taken as the IS and FS, respectively.

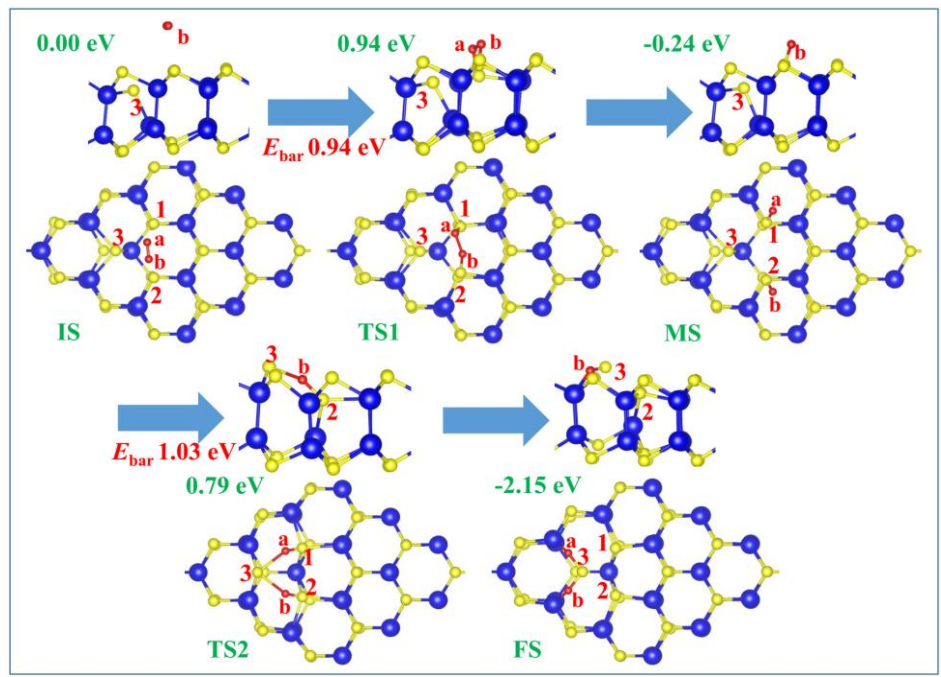

**Figure 10.** The atomic configurations of the IS, TS1, MS, TS2, and FS along the MEP

for the O$_2$ dissociation on the V$_{In}$ surface. The energies of all the states are given with respect to the total energy of the IS. The V$_{In}^1$-PS1(O$_2$) (Fig. 5(a)), V$_{In}$-DC1(O$_2$) (Fig. 5(b)), and V$_{In}$-DC2(O$_2$) (Fig. 5(c)) configurations are taken as the IS, MS, and FS, respectively.

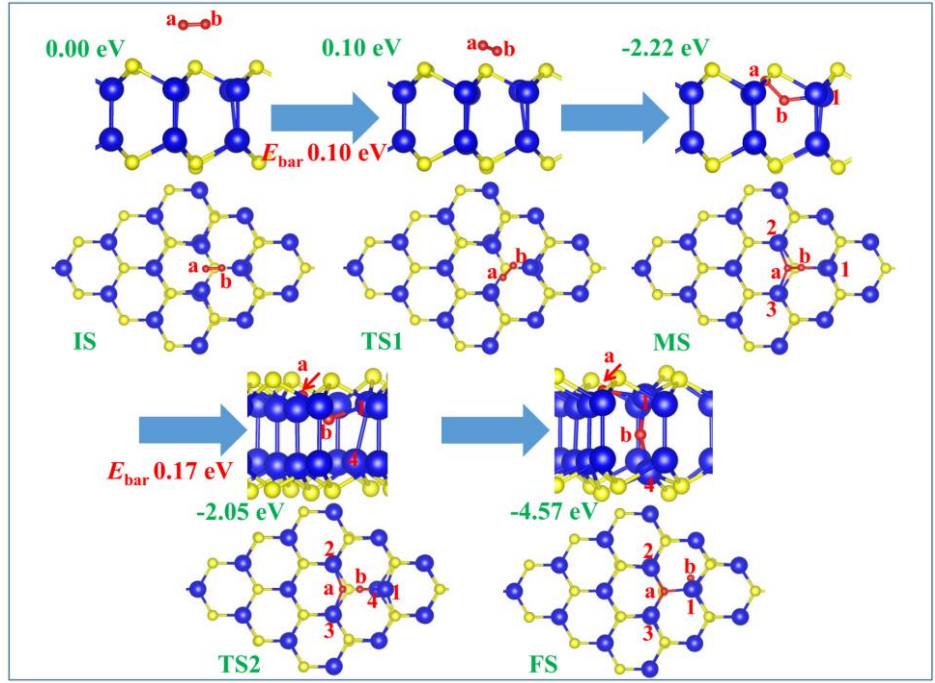

**Figure 11.** The atomic configurations of the IS, TS1, MS, TS2, and FS along the MEP for the O$_2$ dissociation on the V$_{Se}$ surface. The energies of all the states are given with respect to the total energy of the IS. The V$_{Se}$-PS1(O$_2$) (Fig. 7(a)), V$_{Se}$-MC1(O$_2$) (Fig. 7(b)), and V$_{Se}$-DC1(O$_2$) (Fig. 7(d)) configurations are taken as the IS, MS, and FS, respectively.

Considering that an elementary reaction with an energy barrier less than 0.9 eV from the DFT calculation could occur at room temperature easily.[30, 65] Even the exposed under-coordinated Se atoms due to the In vacancy can play an important role in dissociating the adsorbed O$_2$ molecule at ambient conditions. And the exposed under-coordinated In atoms due to the Se vacancy are much more effective for the O$_2$ dissociation compared with the under-coordinated Se atoms. Therefore, the surface oxidation of 2D InSe semiconductors should be dominated by the defects that expose under-coordinated host atoms, especially In atoms. Accordingly, passivation and repairment of the structural defects may be an effective strategy to enhance the chemical stability of 2D InSe semiconductors and thus the device performance.[41, 64, 66] Very recently, Balakrishnan et al. demonstrated that, the InSe thin films exposed to air under ambient conditions are not affected for over 72 hours, while for the same sample, a thermal annealing at 448 $K$ in air can significantly accelerate the oxidation and convert a few surface layers of InSe into $In_2O_3$ over a very short period of 1 hour. The authors suggested that Se tends to desorb from the surface under elevated temperature, leaving behind nucleation sites for O adsorption.[33] This is in line with our present theoretical prediction that the exposed In atoms due to the Se vacancies have extremely high chemical activity towards the adsorbed $O_2$.

## 4. Conclusions

In summary, by using first-principles calculation, the adsorption of $O_2$ and $H_2O$ molecules on the pristine and the defective InSe monolayers is studied. It is predicted that the single Se and In vacancies exhibit significantly enhanced chemical activity toward the adsorbates, and the Se vacancies have a much higher chemical activity than the In vacancies. However, $H_2O$ molecule should be only physisorbed on the various InSe monolayers at ambient conditions, according to the calculated energies. Electronic structure calculation shows that the pristine InSe monolayer and the $V_{In}$ system is *p*-doped by the physisorbed $H_2O$, while the $V_{Se}$ system can be *n*-doped by the physisorbed $H_2O$. Therefore, the physisorbed $H_2O$ will play an important role in modifying the electronic structure of the 2D InSe semiconductor. These results will help us understand the *p*-doing of the InSe reported in Ref.[32].

The vacancies show a much higher chemical activity toward $O_2$ than $H_2O$. Although $O_2$ molecules are still physisorbed on the pristine InSe monolayer, which is *p*-doped by the adsorbates, they will be chemisorbed on the defective InSe monolayers. It is found that the $O_2$ dissociation process is endothermic by 1.24 eV on the pristine InSe monolayer. However, this process is exothermic by 2.15 and 4.57 eV on the $V_{In}$ and $V_{Se}$ surfaces, respectively, Furthermore, the highest barrier for the whole $O_2$ dissociation process is only 0.17 eV for the $V_{Se}$ surface, while they are 2.85 eV for the pristine InSe monolayer and 1.03 eV for the $V_{In}$ system. Therefore, the surface oxidation of 2D InSe semiconductors should be dominated by the defects that expose under-coordinated host atoms, especially In atoms. The above results may give us more insight into the accelerated oxidation of the InSe thin films due to the thermal annealing under elevated temperature in Ref.[33].

Our theoretical results can help better understanding the doping and the oxidation of the 2D InSe semiconductor under ambient conditions. These insights will provide guidance for experimental control and tailing the physical properties of 2D InSe semiconductors, and advance the application of this fascinating material.

**Acknowledgements:**

This work is supported by the Henan Joint Funds of the National Natural Science Foundation of China (Grant No. U1504108), National Natural Science Foundation of China (Grant Nos. 11674083, 11447001, and 21603109). Innovation Scientists and Technicians Troop Construction Projects of Henan Province (No. C20150029).

**Table of Contents Graphic:**

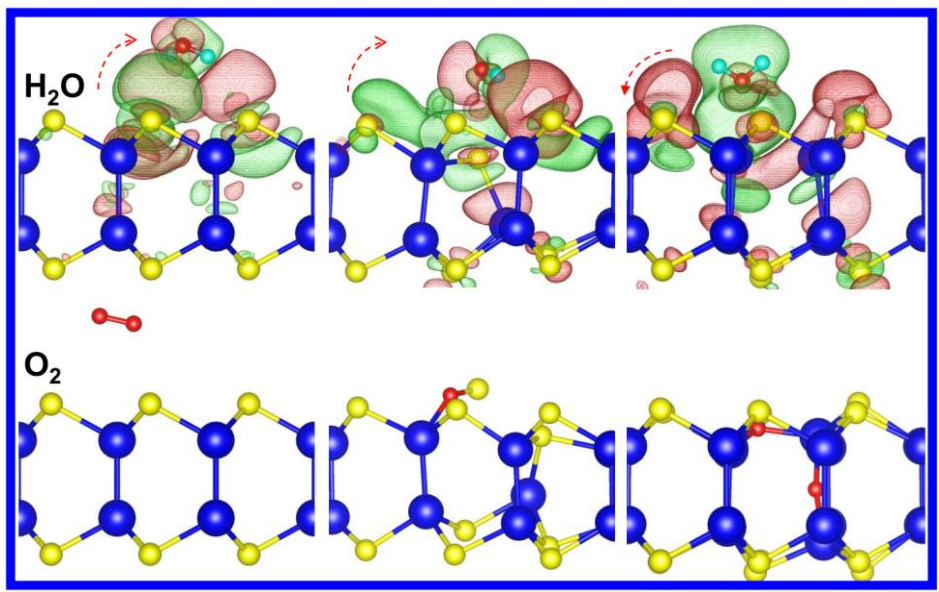